\documentstyle[epsfig,12pt]{article}
\newcommand{\dd}{{\mbox{d}}}
\newcommand{\matrm}[1]{{\mbox{#1}}}
\newcommand{\Li}{{\mbox{Li}_2}}
\newcommand{\vecc}[1]{\mbox{\boldmath $#1$}}

\title{Radiative corrections \\
to the process $\mu^+ \mu^- \to H \gamma$}

\author{A.B.~Arbuzov$^{1}$, E.A.~Kuraev$^{1}$, F.F.~Tikhonin$^{2}$ and \\
B.G. Shaikhatdenov$^{1,3}$}
\date{}

\begin{document}
\maketitle

\begin{center}
{
$^{1}$ \it Joint Institute for Nuclear Research, Dubna, 141980, Russia \\
$^{2}$ \it Institute of High Energy Physics, Protvino, 142284, Russia \\
$^{3}$ \it Institute of Physics and Technology, Almaty, 480082, Kazakstan
} \\[.2cm]
\end{center}

\begin{abstract}
QED radiative corrections to the cross--section
of muon--antimuon annihilation into Higgs boson and photon are
calculated within the 1--loop approximation. We write down the
expression for cross--section in the form of Drell--Yan process, taking
into account higher order leading logs. The non--singlet structure
functions of fermions are shown to obey here evolution
equations of twist--3 operators.
Numerical estimation shows an importance of the correction
in the region close to the threshold of Higgs production.
\end{abstract}

\section{Introduction}

The search for Higgs particles is one of important goals of the
future high--energy colliders~\cite{1}. Most interesting and crucial
parameters are the values of the Higgs couplings to other
fundamental particles. A measurement of the couplings would
allow us to make a choice between different Higgs schemes.

Presumably, Higgs boson may be revealed at the forthcoming LHC collider,
but the precision will be insufficient for the aims outlined above.
Future muon colliders can provide a possibility to study the Higgs
properties in detail.
The idea to collide $\mu\mu$-lepton beams was discussed
long ago~\cite{2}. Its recent development
and a discussion of physics goals are given, for example, in paper~\cite{3}.
Nowadays there are two designs for $\mu \mu$ colliders: \\
1) $\sqrt s = 500\ \matrm{GeV},\qquad {\cal L} = 10^{33}\ cm^{-2} s^{-1},
\qquad {\cal L}_{tot} = 50\ \matrm{fb}^{-1}/\matrm{year}$ ; \\
2) $\sqrt s = 4\ \matrm{TeV},\qquad {\cal L} = 10^{35}\ cm^{-2} s^{-1},
\qquad {\cal L}_{tot}=200 \div 1000\ \matrm{fb}^{-1}/\matrm{year}$.

A $\mu \mu$ collider has definite advantages as compared to $e^+e^-$ one: \\
1) beamstrahlung and bremsstrahlung essentially absent ; \\
2) $m_{\mu} \gg m_{e}$ ; \\
3) focus problems absent, energy resolution $\sim 0.1 \%$,
 small diameter due to absence of synchrotron radiation (cost
 decreases).

At the same time there are also disadvantages: the problem of cooling,
expensive detectors, polarization implies significant loss in ${\cal L}$,
only annihilation channel works $(J = 1)$
hence $\sigma \sim s^{-1}, s \to \infty$.

But apparently, the main positive feature of the $\mu^+ \mu^-$ collider
is the possibility to study the $s$-channel Higgs boson production
({\it Higgs factory}), because the cross--section of the latter
is proportional to the lepton mass.   However, when
$m_H \geq 2M_W$, it is better to study Higgs boson production
in association  with the $Z^0$-boson or photon~\cite{4}. In the
latter case at the Born level the cross--section for the process
\begin{eqnarray}
\mu^-(p_1)\ +\ \mu^+(p_2)\ \longrightarrow\ H(q)\ +\ \gamma(k_1)
\end{eqnarray}
has, at high energies, the following form
\begin{eqnarray}
&& \frac{\dd\sigma^{\gamma H}_0}{\dd c}=\frac{\pi\alpha^2(s^2+M_H^4)}
{2\sin\theta_W^2(1-c^2)s^2(s-M_H^2)}\biggl(\frac{m_\mu}{M_W}\biggr)^2\, ,
\\ \nonumber
&& c=\cos(\widehat{\vecc{p}_1\vecc{k}}_1),
\qquad 1-|c|\gg m_\mu^2/M_H^2.
\end{eqnarray}
Sharp dependence on photon energy $\omega=(s-M_H^2)/(2\sqrt{s})$ in
Eq.(2) makes manifest a reason why the calculations of the radiative
corrections (RC) to this process are urgently desirable, so
we proceed with it. In paper~\cite{Abu} it was shown that in the
region $2\omega/\sqrt{s} \sim 1$ the main contribution
to the cross--section arises from 1--loop electroweak corrections.

\section{Radiative corrections}

The result of the first order RC to the
differential cross--section of the process $\mu^+ \mu^- \to H \gamma$
due to virtual (Fig.~1) and soft real photons
in the case when photon is hitting a non--forward detector
looks as follows:
\begin{eqnarray} \label{bsv}
\frac{\dd\sigma^{\gamma H}_{B+S+V}}{\dd c}&=&\frac{\dd\sigma^{\gamma H}_0}
{\dd c}\left(1+\frac{2\alpha}{\pi}\ln\frac{s}{m^2_\mu}\ln\frac
{\Delta\varepsilon}{\varepsilon}\right)(1-\frac{\alpha}{2\pi}K)\,,
 \\ \nonumber
K&=&-4+2\Li\left(1+\frac{ \chi_1}{M_H^2}\right)
+2\Li\left(1+\frac{ \chi_2}{M_H^2}\right)
-4\Li\left(1-\frac{M_H^2}{ s}\right) - \nonumber\\
&-&\frac{8\pi^2}{3}
-\ln^2\left(\frac{ s}{ \chi_1}\right)
-\ln^2\left(\frac{ s}{ \chi_2}\right)
+\ln^2\left(\frac{M_H^2}{ \chi_1}\right)
+\ln^2\left(\frac{M_H^2}{ \chi_2}\right) - \nonumber\\
&-&\frac{M_H^2( s-M_H^2)+12\chi_1\chi_2\zeta_2}{ s^2
+M_H^4}\,, \\ \nonumber
\Li(z)&=&-\int_0^1\frac{\ln(1-zx)}{x}\dd x, \qquad
\chi_{1,2} = 2p_{1,2}k_1.
\end{eqnarray}
For hard (with energy more than $\Delta\varepsilon$)
collinear photon emission we obtain
\begin{eqnarray} \label{hard}
\frac{\dd\sigma^{hard}}{\dd c}&=& \frac{\alpha}{\pi}
 \int_{\Delta\varepsilon/\varepsilon}^{1}\frac{\dd x}{x}
 \biggl[\left(1-x+\frac{x^2}{2}\right)\ln\frac{s}{m^2_\mu}
-\left(1 - x \right) \biggr] \times \nonumber\\
&\times&\left[ \frac{\dd\widetilde{\sigma}_0(p_1(1-x),p_2)}{\dd c}
+ \frac{\dd\widetilde{\sigma}_0(p_1,p_2(1-x))}{\dd c}\right]\, ,
\end{eqnarray}
where $\dd\widetilde{\sigma}_0((1-x)p_1,p_2)/\dd c$ and
$\dd\widetilde{\sigma}_0(p_1,(1-x)p_2)/\dd c$ are the so--called
{\it shifted} (or {\it boosted}) cross--section. In general case, when
photon emission is allowed from both initial leptons, we have
\begin{eqnarray}
\frac{\dd\widetilde{\sigma}_0(z_1p_1,z_2p_2)}{\dd c}
= \frac{\pi\alpha^2}{2\sin\theta_W^2}\left(\frac{m_\mu}{M_W}\right)^2
\frac{s^2z_1^2z_2^2+M_H^4}{s^2z_1z_2(1-c^2)[sz_1z_2-M_H^2]}\, .
\end{eqnarray}
One can see the cancellation of $\Delta\varepsilon/\varepsilon$
when integrate over $x$ from
$\Delta\varepsilon/\varepsilon$ up to 1.

Summing up the leading terms from Eqs.~(\ref{bsv},\ref{hard}),
we get the first order leading logarithmic correction in the form
\begin{eqnarray} \label{loop1}
\frac{\dd\sigma^{hard+S+V}}{\dd c}
&=& \frac{\alpha}{2\pi} \ln\frac{s}{m^2_\mu}
\int\limits_{0}^{1}\dd x \left(\frac{1}{x}\right)_{+} (1+(1-x)^2)
\biggl[ \frac{\dd\widetilde{\sigma}_0((1-x)p_1,p_2)}{\dd c} + \nonumber\\
&+& \frac{\dd\widetilde{\sigma}_0(p_1,(1-x)p_2)}{\dd c}\biggr]\, .
\end{eqnarray}
The {\it plus operation} acts as usually:
\begin{eqnarray}
\left(\frac{1}{x}\right)_{+} f(x) = \frac{f(x)-f(0)}{x}\, .
\end{eqnarray}

Our result may be compared with that obtained earlier~\cite{5}
on RC to the fermionic width of Higgs in the case $M^2_H \gg m_{\mu}^2$.
Indeed, looking at the leading logarithms we see an agreement.

The contribution of higher orders of perturbation theory can be
taken into account in the leading logarithmic approximation.
Really, as opposite to the case of electron--positron annihilation
into one virtual photon (or $Z$-boson)~\cite{kf,am}
we have to use here another kernel for evolution equations.
Considering only the non--singlet structure functions, we
write down:
\begin{eqnarray}
\widetilde{D}(z,L) &=& \delta(1-z)+ \\ \nonumber
&+& \frac{\alpha}{2\pi}(L-1) \widetilde{P}^{(1)}(z)
+ \frac{1}{2!} \left(\frac{\alpha}{2\pi}(L-1)\right)^2\widetilde{P}^{(2)}(z)
+ \dots\, ,\\ \nonumber
L&=& \ln\frac{s}{m^2_{\mu}},\quad
\widetilde{P}^{(1)}(z)= \lim\limits_{\Delta\to 0}\biggl\{
\frac{1+z^2}{1-z}\Theta(1-z-\Delta)+ \\ \nonumber
&+& 2\delta(1-z)\ln\Delta \biggr\}, \qquad
\int\limits_{0}^{1}\dd z \widetilde{P}^{(1)}(z)=- \frac{3}{2}\, ,\\ \nonumber
\widetilde{P}^{(2)}(z) &=& \int\limits_{z}^{1}\frac{\dd t}{t}
\widetilde{P}^{(1)}(t)\widetilde{P}^{(1)}\biggl(\frac{z}{t}\biggr).
\end{eqnarray}
A {\it smoothed} representation (analogous to the one derived
in Ref.~\cite{kf}) for the modified D--function looks as follows:
\begin{eqnarray}
\widetilde{D}(z,L) &=& - \frac{\dd}{\dd z}\int\limits_{z}^{1}\dd x \,
\widetilde{D}(x,L) = \frac{1}{2}\beta (1-z)^{\beta/2-1}(1+z^2)
(1 + {\cal O}(\beta^2)), \\ \nonumber
\beta &=& \frac{2\alpha}{\pi}(L-1).
\end{eqnarray}

The master formula for radiatively corrected cross--section
has the form of the Drell--Yan cross--section. So, we suggest
to write the result as a convolution of the modified
lepton structure functions with the shifted cross--section of
the hard subprocess. It reads
\begin{eqnarray}
\frac{\dd\sigma}{\dd c} &=& \!\!\int\limits_{z_1^{min}}^{1}\dd z_1
\widetilde{D}(z_1) \!\!
\int\limits_{z_2^{min}}^{1}\dd z_2 \widetilde{D}(z_2)
\frac{\dd\widetilde{\sigma}_0(z_1p_1,z_2p_2)}{\dd c}
\biggl(1 -
\frac{\alpha}{2\pi}K\biggr)\Theta(\omega-\omega_{th}),
\end{eqnarray}
where a part of non--leading terms is taken into account by the
$K$--factor and $\omega_{th}$ is the experimental energy threshold
of the photon registration.
The energy conservation law gives us the energy of the detected photon
\begin{eqnarray}
\omega = \frac{sz_1z_2-M_H^2}{2\varepsilon(z_1+z_2-c(z_1-z_2))}\, .
\end{eqnarray}
The lower limits for integration over $z_{1,2}$ are to be defined also
just from the above expression by imposing the condition
$\omega >\omega_{th}$:
\begin{eqnarray}
z_1^{min} = \frac{M_H^2 + \sqrt{s}\omega_{th}(1+c)}
{s-\sqrt{s}\omega_{th}(1-c)}\, ,\qquad
z_2^{min} = \frac{M_H^2 + \sqrt{s}z_1\omega_{th}(1-c)}
{sz_1-\sqrt{s}\omega_{th}(1+c)}\, .
\end{eqnarray}

\begin{figure}
\unitlength=0.65mm
\special{em:linewidth 0.4pt}
\linethickness{0.4pt}
\begin{picture}(222.00,66.33)
\put(123.14,61.00){\line(4,-5){20.00}}
\put(125.80,57.67){\vector(3,-4){1.67}}
\put(138.80,41.67){\vector(3,-4){1.67}}
\put(134.81,28.00){\oval(2.00,2.00)[l]}
\put(134.81,30.00){\oval(2.00,2.00)[r]}
\put(134.81,32.00){\oval(2.00,2.00)[l]}
\put(134.81,34.00){\oval(2.00,2.00)[r]}
\put(134.81,36.00){\oval(2.00,2.00)[l]}
\put(134.81,38.00){\oval(2.00,2.00)[r]}
\put(134.81,40.00){\oval(2.00,2.00)[l]}
\put(134.81,42.00){\oval(2.00,2.00)[r]}
\put(134.81,44.00){\oval(2.00,2.00)[l]}
\put(179.81,61.00){\line(4,-5){20.00}}
\put(194.81,43.66){\oval(2.00,3.33)[lt]}
\put(194.81,47.00){\oval(2.00,3.33)[rb]}
\put(196.81,46.66){\oval(2.00,3.33)[lt]}
\put(196.81,50.00){\oval(2.00,3.33)[rb]}
\put(198.81,49.66){\oval(2.00,3.33)[lt]}
\put(198.81,53.00){\oval(2.00,3.33)[rb]}
\put(200.81,52.66){\oval(2.00,3.33)[lt]}
\put(200.81,56.00){\oval(2.00,3.33)[rb]}
\put(202.81,55.66){\oval(2.00,3.33)[lt]}
\put(202.81,59.00){\oval(2.00,3.33)[rb]}
\put(204.81,58.66){\oval(2.00,3.33)[lt]}
\put(204.81,62.00){\oval(2.00,3.33)[rb]}
\
\put(182.47,57.67){\vector(3,-4){1.67}}
\put(189.14,49.67){\vector(3,-4){1.67}}
\put(195.47,41.67){\vector(3,-4){1.67}}
\put(131.80,51.33){\oval(2.00,3.33)[lt]}
\put(131.80,54.67){\oval(2.00,3.33)[rb]}
\put(133.80,54.33){\oval(2.00,3.33)[lt]}
\put(133.80,57.67){\oval(2.00,3.33)[rb]}
\put(135.80,57.33){\oval(2.00,3.33)[lt]}
\put(135.80,60.67){\oval(2.00,3.33)[rb]}
\put(137.80,60.33){\oval(2.00,3.33)[lt]}
\put(130.80,51.67){\vector(3,-4){1.67}}
\
\put(14.00,61.00){\line(4,-5){20.00}}
\put(14.00,11.00){\line(4,5){20.00}}
\put(21.00,53.51){\oval(2.00,2.00)[t]}
\put(23.00,53.51){\oval(2.00,2.00)[b]}
\put(25.00,53.51){\oval(2.00,2.00)[t]}
\put(27.00,53.51){\oval(2.00,2.00)[b]}
\put(29.00,53.51){\oval(2.00,2.00)[t]}
\put(31.00,53.51){\oval(2.00,2.00)[lb]}
\put(31.00,51.51){\oval(2.00,2.00)[r]}
\put(31.00,49.51){\oval(2.00,2.00)[l]}
\put(31.00,47.51){\oval(2.00,2.00)[r]}
\put(31.00,45.51){\oval(2.00,2.00)[l]}
\put(31.00,43.51){\oval(2.00,2.00)[r]}
\put(31.00,41.51){\oval(3.00,2.00)[lt]}
\put(16.00,58.50){\vector(3,-4){1.67}}
\put(24.00,48.50){\vector(3,-4){1.67}}
\put(30.00,41.00){\vector(3,-4){1.67}}
\put(19.80,55.33){\oval(2.00,3.33)[lt]}
\put(19.80,58.67){\oval(2.00,3.33)[rb]}
\put(21.80,58.33){\oval(2.00,3.33)[lt]}
\put(21.80,61.67){\oval(2.00,3.33)[rb]}
\put(23.80,61.33){\oval(2.00,3.33)[lt]}
\put(23.80,64.67){\oval(2.00,3.33)[rb]}
\put(29.00,62.00){\makebox(0,0)[cc]{$\gamma$}}
\put(10.14,59.00){\makebox(0,0)[cc]{$\mu$}}
\
\put(69.00,61.00){\line(4,-5){20.00}}
\put(76.00,53.51){\oval(2.00,2.00)[t]}
\put(78.00,53.51){\oval(2.00,2.00)[b]}
\put(80.00,53.51){\oval(2.00,2.00)[t]}
\put(82.00,53.51){\oval(2.00,2.00)[b]}
\put(84.00,53.51){\oval(2.00,2.00)[t]}
\put(86.00,53.51){\oval(2.00,2.00)[lb]}
\put(86.00,51.51){\oval(2.00,2.00)[r]}
\put(86.00,49.51){\oval(2.00,2.00)[l]}
\put(86.00,47.51){\oval(2.00,2.00)[r]}
\put(86.00,45.51){\oval(2.00,2.00)[l]}
\put(86.00,43.51){\oval(2.00,2.00)[r]}
\put(86.00,41.51){\oval(3.00,2.00)[lt]}
\put(71.00,58.50){\vector(3,-4){1.67}}
\put(85.80,40.33){\vector(3,-4){1.67}}
\put(82.00,46.00){\oval(2.00,3.00)[lt]}
\put(82.00,49.00){\oval(2.00,3.00)[rb]}
\put(84.00,49.00){\oval(2.00,3.00)[lt]}
\put(84.00,52.00){\oval(2.00,3.00)[rb]}
\put(86.00,52.00){\oval(2.00,3.00)[lt]}
\put(86.00,55.00){\oval(2.00,3.00)[rb]}
\put(88.00,55.00){\oval(2.00,3.00)[lt]}
\put(88.00,58.00){\oval(2.00,3.00)[rb]}
\put(90.00,58.00){\oval(2.00,3.00)[lt]}
\put(90.00,61.00){\oval(2.00,3.00)[rb]}
\put(86.00,62.00){\makebox(0,0)[cc]{$k_1$}}
\put(65.00,59.00){\makebox(0,0)[cc]{$p_1$}}
\put(13.00,17.00){\makebox(0,0)[cc]{$\bar \mu$}}
\put(69.00,11.00){\line(4,5){20.00}}
\put(65.00,17.00){\makebox(0,0)[cc]{$-p_2$}}
\put(39.00,22.00){\makebox(0,0)[cc]{(1)}}
\put(96.00,22.00){\makebox(0,0)[cc]{(2)}}
\put(123.00,11.00){\line(4,5){20.00}}
\put(180.00,11.00){\line(4,5){20.00}}
\put(92.00,46.00){\makebox(0,0)[cc]{$k$}}
\put(37.00,46.00){\makebox(0,0)[cc]{$k$}}
\put(51.00,41.00){\makebox(0,0)[cc]{$H$}}
\put(109.00,41.00){\makebox(0,0)[cc]{$q$}}
\put(34.00,36.00){\line(1,0){22.00}}
\put(56.00,36.00){\line(0,1){0.00}}
\put(89.00,36.00){\line(1,0){22.00}}
\put(143.00,36.00){\line(1,0){22.00}}
\put(200.00,36.00){\line(1,0){22.00}}
\put(188.81,28.00){\oval(2.00,2.00)[l]}
\put(188.81,30.00){\oval(2.00,2.00)[r]}
\put(188.81,32.00){\oval(2.00,2.00)[l]}
\put(188.81,34.00){\oval(2.00,2.00)[r]}
\put(188.81,36.00){\oval(2.00,2.00)[l]}
\put(188.81,38.00){\oval(2.00,2.00)[r]}
\put(188.81,40.00){\oval(2.00,2.00)[l]}
\put(188.81,42.00){\oval(2.00,2.00)[r]}
\put(188.81,44.00){\oval(2.00,2.00)[l]}
\put(188.81,46.00){\oval(2.00,2.00)[r]}
\put(188.81,48.00){\oval(2.00,2.00)[l]}
\put(188.81,26.00){\oval(2.00,2.00)[r]}
\put(188.81,24.00){\oval(2.00,2.00)[l]}
\put(150.00,22.00){\makebox(0,0)[cc]{(3)}}
\put(208.00,22.00){\makebox(0,0)[cc]{(4)}}
\put(26.00,26.00){\vector(-1,-1){1.00}}
\put(81.00,26.00){\vector(-1,-1){1.00}}
\put(128.00,17.00){\vector(-1,-1){0.97}}
\put(185.00,17.00){\vector(-1,-1){0.97}}
\end{picture}
\caption{ {\small The subset of Feynman diagrams for the process
$\mu \bar\mu \to H\gamma$.}}
\end{figure}
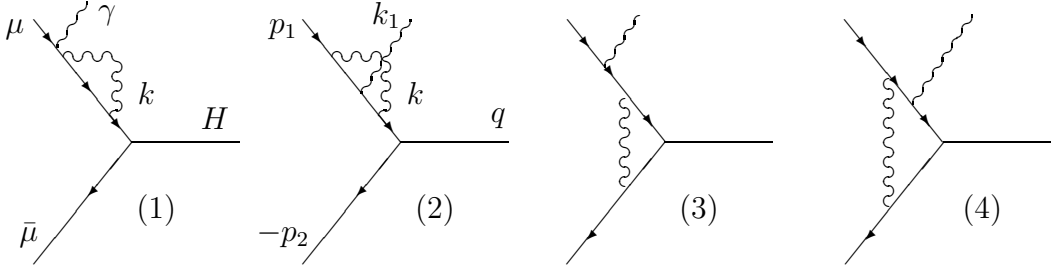

\section{Conclusions}

We use the normalization for $e\bar eH(q)$ vertex
function $\Gamma^{(2)}(q^2)|_{q^2=0}=0$ as is akin to one exploited
for $\, e\bar eZ\ $vertex in the paper of Berends et al.~\cite{7}.
The choice of subtraction point $q^2=0$ is common to the Standard
Model parameters $\sin\theta_W$, $M_Z$, $M_H$ normalization.

\begin{figure}[t]
\begin{center}
\mbox{\epsfig{file=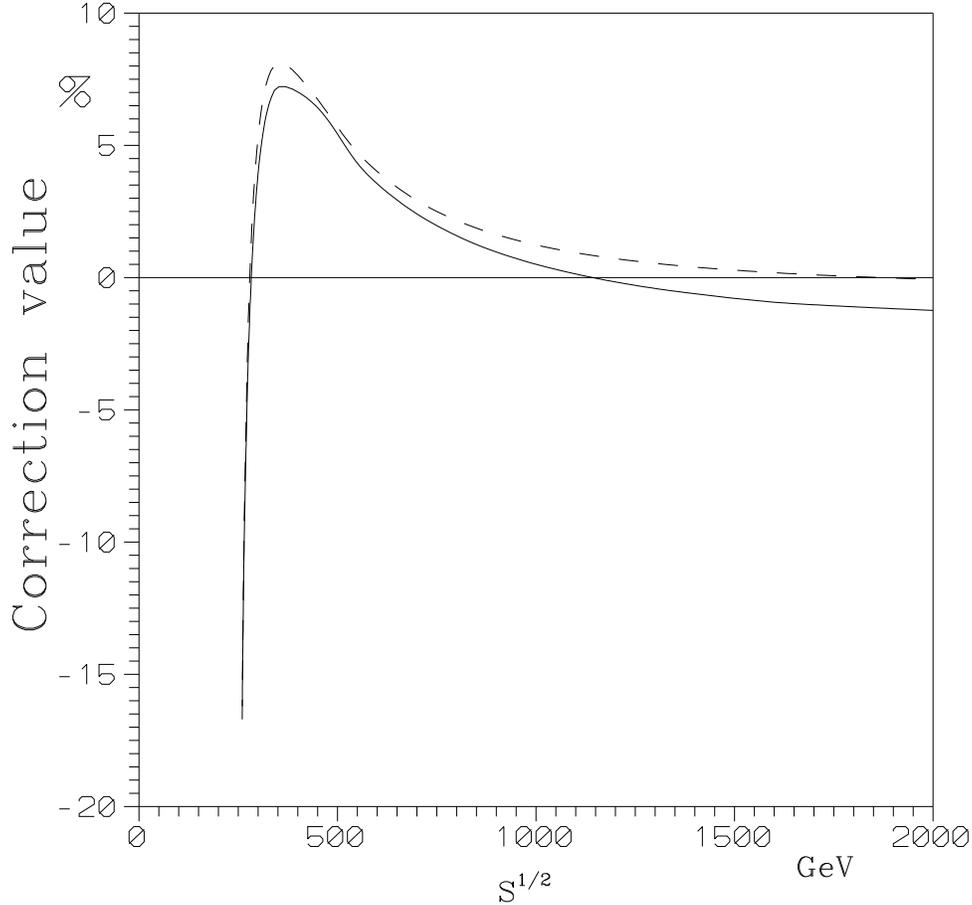,height=12cm,angle=0}}
\end{center}
\caption{ \small
Radiative corrections to the process $\mu \bar\mu \to H\gamma$. }
\end{figure}

The kernel of the evolution equation for $\tilde D(z)\ ,
\tilde P^{(1)}(z)$ differs from the one,
that appears in the evolution equation for $D^{NS}(z)$~\cite{8}.
The kernel $P^{(1)}(z)$ is responsible for a, say, single photon annihilation
of $\mu^+ \mu^-$ into hadrons or the leading twist contribution
in deep inelastic scattering.
This fact is natural, since the kernels $\widetilde{P}^{(1)}(z)$ and
$P^{(1)}(z)$  describe evolution of matrix elements of twist--3
operators $\bar\psi \psi$ and twist--2 operators
$\ \bar\psi \gamma_{\mu} \psi$, respectively.

Due to the fact that the cross--section of
$\mu^+ \mu^- \to H \gamma$ is proportional to muon mass squared,
the validity of the Kinoshita--Lee--Nauenberg theorem~\cite{9}
proving the absence of singularities in the limit $m_{\mu}\to 0$
is restored. The problem was first noted and discussed
in calculations of radiative corrections to the Higgs
decay width into fermions~\cite{10}.

In the Fig.~2 we presented the values of radiative corrections
as functions of the center--of--mass energy
\begin{eqnarray}
\delta(\sqrt{s}) = \frac{ \int\limits_{c_{\mathrm{min}}}^{c_{\mathrm{max}}}\dd c
(\dd\sigma/\dd c)}{\int\limits_{c_{\mathrm{min}}}^{c_{\mathrm{max}}}\dd c
(\dd\sigma_0^{\gamma H}/\dd c)}\, 100\%.
\end{eqnarray}
We took $M_H=250$~GeV,
the value of photon energy threshold $\omega_{th}$=5~GeV, and
the angular range for photon detection $-0.999 < c < 0.999$.
The dashed line represents the first order leading logarithmic
correction, calculated according to Eq.~(\ref{loop1})
with $K$-factor included.
The solid line shows the values of the complete RC according to Eq.~(11).

\subsection*{Acknowledgments}
We are grateful to V.S.~Fadin and L.N.~Lipatov for discussions.
Two of us (A.A., E.K.) are grateful to the INTAS
foundation, grant 93--1867 ext.

\end{document}